\documentclass[manuscript,noacm]{acmart}

\usepackage{amsmath,amsfonts}
\usepackage{mathtools,xparse}
\usepackage{pifont}
\usepackage{graphicx}

\usepackage{setspace}
\usepackage{booktabs}

\usepackage{epstopdf}
\usepackage{xspace}

\usepackage{algorithm}
\usepackage{algpseudocode}
\algdef{SE}[DOWHILE]{Do}{doWhile}{\algorithmicdo}[1]{\algorithmicwhile\ #1}%

\usepackage{cprotect}
\usepackage{soul}
\usepackage{caption}
\usepackage{subcaption}
\usepackage{tablefootnote}

\usepackage{makecell}
\usepackage{multirow}
\usepackage{multicol,tabularx,capt-of}
\usepackage{fancyvrb}
\usepackage{hyperref}

\usepackage{enumitem}

\usepackage{tikz}
\usepackage{xcolor}
\newcommand*\circled[1]{\tikz[baseline=(char.base)]{
            \node[shape=circle,fill,inner sep=1.0pt] (char) {\textcolor{white}{\small#1}};}}

\newcommand{\llama}{Llama2\xspace}

\newcommand{\reffigure}{Fig.\xspace}
\newcommand{\reftable}{Table\xspace}

\newcommand{\Design}{SLIM\xspace}
\newcommand{\Designcap}{SLIM\xspace}

\newcommand{\ie}{\textit{i}.\textit{e}.\xspace}
\newcommand{\eg}{\textit{e}.\textit{g}.\xspace}

\AtBeginDocument{%
  }

\acmConference[Conference acronym 'XX]{Make sure to enter the correct
  conference title from your rights confirmation emai}{June 03--05,
  2018}{Woodstock, NY}
\acmISBN{978-1-4503-XXXX-X/18/06}

\begin{document}

\title[\small{\Design: A Heterogeneous Accelerator for Edge Inference of Sparse Large Language Model via Adaptive Thresholding}]{\Design: A Heterogeneous Accelerator for Edge Inference of Sparse Large Language Model via Adaptive Thresholding}

\author{Weihong Xu}
\email{wexu@ucsd.edu}
\affiliation{%
  \institution{Department of Computer Science and Engineering, University of California San Diego}
  \city{La Jolla}
  \state{California}
  \country{USA}
}

\author{Haein Choi}
\email{hac046@ucsd.edu}
\affiliation{%
  \institution{Department of Computer Science and Engineering, University of California San Diego}
  \city{La Jolla}
  \state{California}
  \country{USA}
}

\author{Po-kai Hsu}
\email{pokai.hsu@gatech.edu}
\affiliation{%
  \institution{School of Electrical and Computer Engineering, Georgia Institute of Technology}
  \city{Atlanta}
  \state{California}
  \country{USA}
}

\author{Shimeng Yu}
\email{shimeng.yu@ece.gatech.edu}
\affiliation{%
  \institution{School of Electrical and Computer Engineering, Georgia Institute of Technology}
  \city{Atlanta}
  \state{California}
  \country{USA}
}

\author{Tajana Rosing}
\email{tajana@ucsd.edu}
\affiliation{%
  \institution{Department of Computer Science and Engineering, University of California San Diego}
  \city{La Jolla}
  \state{California}
  \country{USA}
}


\renewcommand{\shortauthors}{Weihong et al.}

\begin{abstract}
    Large language models (LLMs), composed of Transformer decoders, have demonstrated unparalleled proficiency in understanding and generating human language. However, efficient LLM inference on resource-constraint embedded devices remains a challenge because of the sheer model size and memory-intensive operations that arise from feedforward network (FFN) and multi-head attention (MHA) layers. 
    Existing accelerations offload LLM inference to heterogeneous computing systems comprising expensive memory and processing units. However, recent studies show that most hardware resources are not used because LLM exhibits significant sparsity during inference. The sparsity of LLMs provides a good opportunity to perform memory-efficient inference.
    In this work, we propose \Design, an algorithm and hardware co-design optimized for sparse LLM serving on the edge. \Design exploits LLM's sparsity by only fetching activated neurons to significantly reduce data movement. To this end, the efficient inference algorithm based on adaptive thresholding is proposed to support runtime configurable sparsity at the cost of negligible accuracy loss. 
    Then, we present the \Design heterogeneous hardware architecture that combines the best of both near-storage processing (NSP) and processing-in-memory (PIM). \Design stores FFN weights in high-density 3D NAND and computes FFN layers in NSP units, alleviating high memory requirements caused by FFN weights. The memory-intensive MHA with low arithmetic density is processed in the PIM module.
    By leveraging the inherent sparsity observed in LLM operations and integrating NSP with PIM techniques within SSDs, \Design significantly reduces memory footprint, data movement, and energy consumption. 
    Meanwhile, we present the software support for integrating design into existing SSD system. Our comprehensive analysis and system-level optimization demonstrate the effectiveness of our sparsity-tailored accelerator, offering 13-18$\times$ throughput improvements over SSD-GPU system and 9-10$\times$ better energy efficiency over DRAM-GPU system while maintaining low latency. 
\end{abstract}




\maketitle

\section{Introduction}

Large language models (LLMs) have become transformative tools in a wide range of applications, including natural language understanding~\cite{wang2023large,karanikolas2023large}, content generation~\cite{kasneci2023chatgpt,sarsa2022automatic,swanson2021story}, and healthcare~\cite{arora2023promise,cascella2023evaluating}. 
Recently, local and edge deployments of LLM inference are becoming increasingly important as they offer significant advantages in cost efficiency, latency reduction, and data privacy over centralized solutions. 
Performing inference directly on edge devices eliminates the need for network connections, reducing reliance on cloud infrastructure and thereby lowering operational costs. Additionally, edge inference minimizes latency by processing data closer to the source. This allows for faster response for real-time applications in fields like autonomous systems and healthcare. Moreover, as data privacy concerns grow, edge LLM inference provides a secure solution by keeping sensitive information on the device, reducing the risks associated with transmitting data over external networks. 
These benefits make edge-based LLM inference not only practical but essential for scalable and secure AI integration across various industries.

LLMs' remarkable capabilities in generating high-quality content are primarily attributed to the scaling of model sizes to billions of parameters, which enables them to store and process vast amounts of knowledge across diverse domains. The model scaling not only improves linguistic fluency and contextual understanding but also enhances the models' ability to handle complex, multi-step reasoning and nuanced information across specialized fields. 
The sheer size of LLM presents significant computational challenges for low-latency and low-cost inference during the deployment phase, because LLM inference demands immense hardware resources and prolonged inference times. 
First, LLM storage exhibits a prohibitive memory footprint. State-of-the-art (SoA) LLMs~\cite{touvron2023llama,zhang2022opt,mixtral,qw_moe} consume over 10GB to 100GB memory space to store the billion-scale parameters. The required memory easily exceeds the capacity of commodity hardware devices, such as graphics processing unit (GPU). 
Second, LLM inference is bottlenecked by the sequential and memory-bound operations from multi-head attention (MHA) and feedforward network (FFN) with low arithmetic intensity~\cite{park2024attacc,heo2024neupims}. LLM's expensive memory footprint and sequential execution not only increase the overall inference latency, but also limit the achievable throughput.

Existing LLM inference is typically served online using server-grade GPUs and high-bandwidth memories. The average inference cost and latency are amortized by allocating large-volume user requests to large-scale hardware facilities~\cite{miao2024spotserve,park2024attacc,heo2024neupims} optimized for LLM, thereby delivering satisfactory service quality and throughput. 
However, the aforementioned solutions relying on high-performance GPUs and memories at scale are not viable for the edge deployment of LLM. This is because edge devices with limited resources are sensitive to deployment cost and hardware complexity. The expensive facilities are difficult to accommodate on edge scenarios.

Consequently, we need a more accessible and cost-effective edge deployment solution of LLMs. To this end, SoA designs perform joint optimizations for the algorithm and hardware components. 
For the algorithm, recent studies~\cite{liu2023dejavu,song2024prosparse,song2023powerinfer} show that LLM inference exhibits significant redundancy. For example, LLM inference exhibits a high degree of sparsity, leading to wasted memory access as substantial data movement involves inactive components. Zero skipping~\cite{liu2023dejavu,mirzadeh2023relustrikeback,song2023powerinfer,hwang2023pregate,song2024prosparse} and model pruning~\cite{wang2024svd,yuan2023asvd,yu2023compressing,chen2021drone,hsu2207language} are the two commonly adopted algorithm optimizations to reduce memory footprint and energy consumption. 
But these methods show severe accuracy loss, limited flexibility, or high fine-tuning cost (see Section~\ref{sec:motivation}). 
From the hardware perspective, SoA LLM accelerators~\cite{park2024attacc,heo2024neupims,sheng2023flexgen,song2023powerinfer,alizadeh2023llminaflash,hwang2023pregate,huang2025edgellm} exploit heterogeneous computing to mitigate the high-cost hardware by offloading different LLM modules to cheaper hardware, {including DRAM/SSDs combined with GPUs~\cite{park2024attacc,heo2024neupims,sheng2023flexgen,song2023powerinfer,alizadeh2023llminaflash,hwang2023pregate} or CPU-FPGA system~\cite{huang2025edgellm}}. The heterogeneous designs effectively handle the large model size and memory-intensive operations. 
Nevertheless, according to our analysis in Section~\ref{sec:motivation}, these designs suffer from slow cold inference~\cite{yi2023cold} and low throughput due to the limited PCIe bandwidth between SSD/DRAM and GPU. Hence, this necessitates a cost-effective LLM solution with improved latency.

In this work, we mainly aim at developing a cost-effective solution for local and edge deployments of LLMs. These application scenarios pose distinct challenges, primarily due to small batch sizes and limited hardware resources. The characteristics of LLM inference under the local or edge environments can be summarized as:
\begin{itemize}
    \item \textbf{Small batch size.} It is difficult for local applications to collect large batch sizes without increasing latency because these scenarios show much lower statistical multiplexing and request volume compared to data center. Therefore, inference with small batch sizes is dominated by the data movement due to the low arithmetic intensity~\cite{heo2024neupims,park2024attacc} because general matrix multiply (GEMM) operations exhibit low weight reuse.
    \item \textbf{Limited hardware resources.} Embedded devices typically provide limited memory space and bandwidth due to their limited hardware resources. This is especially challenging because single device is unable to provide sufficient memory to store the entire LLM. Meanwhile, the low bandwidth significantly increases LLM inference latency. 
\end{itemize}

To address these challenges, we propose a algorithm-hardware co-design, \Design, which is a DRAM-SSD heterogeneous accelerator optimized for serving sparse LLM inference. First, \Design has an efficient inference mechanism with runtime configurable sparsity. Second, \Design heterogeneous hardware architecture combines the high-density storage capabilities of SSD and processing-in-memory (PIM) capabilities from DRAM. The heterogeneous design enjoys the benefits of SSD and DRAM-PIM, thereby improving computational efficiency by selectively processing activated neurons. To summarize, our contributions are as follows:
\begin{enumerate}
    \item \textbf{Efficient inference with adaptive sparsity.} \Design utilizes a low-rank sparsity prediction with thresholding to effectively realize tradeoff between sparsity and inference speed, significantly reducing redundant data movement and computations. This approach not only maximizes efficiency by processing only the activated neurons, but also significantly reduces the power consumption and latency associated with LLM inference. Compared to other methods~\cite{alizadeh2023llminaflash,song2023powerinfer,wang2024svd}, our design shows superior advantages in terms of training cost, accuracy loss, and runtime flexibility.
    \item \textbf{Heterogeneous accelerator architecture.} \Design introduces a novel heterogeneous accelerator design that integrates NSP-based SSD and PIM-based DRAM. This architecture is specifically optimized for sparse LLM serving. By integrating the inference into SSD-DRAM system, \Design offers a more cost-effective solution compared to existing GPU-centric solutions~\cite{song2023powerinfer,alizadeh2023llminaflash}. This makes it particularly suitable for deployment in resource-constrained environments, where both economic and physical space constraints often exist. Meanwhile, we also present various schemes, including data mapping, execution scheduling, and interface design, to optimize the hardware utilization and system efficiency. 
    \item \textbf{Evaluation.} Our evaluation demonstrates that \Design achieves up to 13-18$\times$ throughput improvement over the SSD-GPU design~\cite{sheng2023flexgen} and 9-10$\times$ energy improvements over the DRAM-GPU system~\cite{song2023powerinfer}. This shows \Design's capability to handle edge LLM inference more effectively than existing solutions.
\end{enumerate}

The remaining sections of this article are organized as follows. Section~\ref{sec:background} introduces the basics for LLMs, SSD, and near-storage processing. Section~\ref{sec:motivation} presents the related work as well as motivational analysis for our proposed design. Section~\ref{sec:design_algorithm} proposes the efficient LLM inference with predicted sparsity and thresholding. Section~\ref{sec:design_arch_sys} introduces the heterogeneous DRAM-SSD hardware architecture that implements the proposed LLM inference. Section~\ref{sec:design_software} presents the software optimization schemes and modifications to integrate the proposed design. Section~\ref{sec:eval} presents the evaluation methodology, experimental results, and analysis. Section~\ref{sec:conclusion} concludes the paper.

\section{Background}\label{sec:background}

\subsection{Large Language Models}
\begin{figure}[t]
    \centering
    \includegraphics[width=0.7\linewidth]{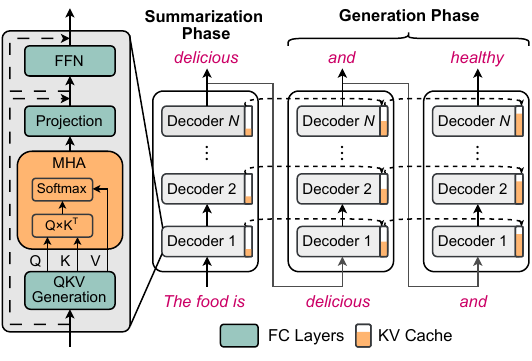}
    \caption{Decoder-based LLM inference.}
    \label{fig:llm_decoder_model}
\end{figure}

\noindent\textbf{Transformer Decoder.} 
\reffigure\ref{fig:llm_decoder_model} illustrates the LLM structure, which is comprised of $N_{dec}$ layers of decoders. Each decoder is constructed from QKV generation, multi-head attention (MHA), coupled with a feed-forward network (FFN) or mixture of experts (MoE) module. Initially, a batch of $L$ input tokens are transformed into their respective $L\times dim_e$-dimensional embedding vectors. Subsequently, these vectors are processed through the decoder module, which is sequentially linked by $N_{dec}$ Transformer decoders~\cite{vaswani2017attention}.

\begin{figure}[t]
    \centering
    \includegraphics[width=0.6\linewidth]{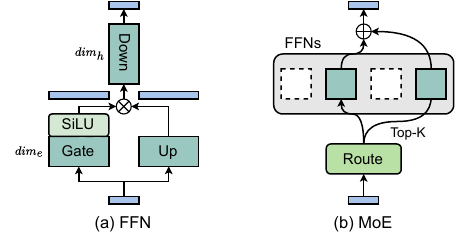}
    \caption{FFN and MoE layers with SiLU activation.}
    \label{fig:llm_ffn_moe}
\end{figure}

First, the QKV generation module produces a $L \times dim_e$ query matrix $\mathbf{Q}$, a $L \times dim_e$ key matrix $\mathbf{K}$, and a $L \times dim_e$ value matrix $\mathbf{V}$, by multiplying the embedding vectors with three weight matrices, respectively. Then $\mathbf{Q}$, $\mathbf{K}$ and $\mathbf{V}$ matrices are fed into the MHA module which computes the scaled dot-product attention as follows:
\begin{equation}\label{eq:mha}
\begin{aligned}
    \mathbf{H}^{(i)} &= \text{Softmax}\left( \frac{\mathbf{Q}^{(i)} \cdot \mathbf{K}^{(i)T}}{\sqrt{dim_e}} \right) \cdot \mathbf{V}^{(i)},\\
    \text{MHA}(\mathbf{X}) &= \text{Concat}\left( \mathbf{H}^{(1)}, \ldots, \mathbf{H}^{(h)} \right) \cdot \mathbf{W}_{o},
\end{aligned}
\end{equation}
where $\text{Softmax}(\cdot)$ denotes the Softmax function. The MHA output is then processed by FFN to generate the input for the next decoder. Most LLMs integrate FFN with a gating mechanism in \reffigure\ref{fig:llm_ffn_moe}-(a) to enhance their ability to modulate the information flow. The gated FFN computation contains two general matrix multiplies (GEMMs) and one element-wise vector multiplication as follows:
\begin{equation}\label{eq:ffn}
    \text{FFN}(\mathbf{X}) =  \Bigl( \sigma(\mathbf{X} \cdot \mathbf{W}_g^T) \odot (\mathbf{X} \cdot \mathbf{W}_{u}^T) \Bigl) \cdot \mathbf{W}_{d}^T,
\end{equation}
where $\odot$ denotes element-wise multiplication while $\sigma$ is the activation. The FFN input is first projected into a hidden $d_h$-dimensional vector using gate weight $\mathbf{W}_g$ and up projection weight $\mathbf{W}_u$. The point-wise result is projected back into original $dim_e$-dimension vector by down projection weight $\mathbf{W}_d$.

\noindent\textbf{Mixture of Experts.} 
Recent LLMs~\cite{mixtral} leverage MoE to achieve scalable performance while maintaining low inference complexity. Specifically, MoE combines multiple FFNs that work as the `experts', and each expert is trained to to specialize in certain subsets of tasks. 
As shown in \reffigure\ref{fig:llm_ffn_moe}-(b), MoE involves a routing process that determines which experts are best suited. Specifically, only the top-$k$ of all $N_{\text{expert}}$ FFNs will be activated to make a final prediction. The MoE route module determines the weight for each expert's prediction based on the input. 
MoEs have been shown to be promising for enabling higher efficiency and accuracy because partially activated FFNs lead to low inference overhead. For example, Mixtral-$8\times7$B~\cite{mixtral} has the comparable amount of FFN parameters as \llama-70B. But it delivers an inference speedup of $6\times$ over \llama-70B since only $25\%$ of all FFNs are activated.

\noindent\textbf{Inference Process.} 
\reffigure\ref{fig:llm_decoder_model} illustrates the inference process of LLMs, which consists of 1. summarization phase and 2. generation phase. These two phases execute identical operations but with different sequence lengths:
\begin{enumerate}
    \item \textbf{Summarization Phase} involves understanding the context of input sequences by processing $L>1$ tokens in parallel. The critical step is called \textit{KV cache prefilling}. The generated key $\mathbf{Q}$ and value $\mathbf{V}$ matrices in MHA module are cached to memorize the context and knowledge of given input. The cached $2\times L \times dim_e$ KV data with will later be reused and updated during generation phase.
    \item \textbf{Generation Phase} uses LLM's autoregressive capabilities to sequentially generate new content. The previously generated token is considered the input of the next time step. Due to data dependency, the generation phase only processes $L=1$ token per time step. In MHA module, the newly generated KV vectors are appended to the cached KV to update knowledge data.
\end{enumerate}

The generation phase is the major bottleneck of LLM inference, incurring $85\%$ to $96\%$ runtime overhead~\cite{park2024attacc}. This is mainly due to its sequential nature and low arithmetic intensity. Hence, we mainly focus on the acceleration for LLM generation in the following work.

\begin{figure}[t]
    \centering
    \includegraphics[width=0.7\linewidth]{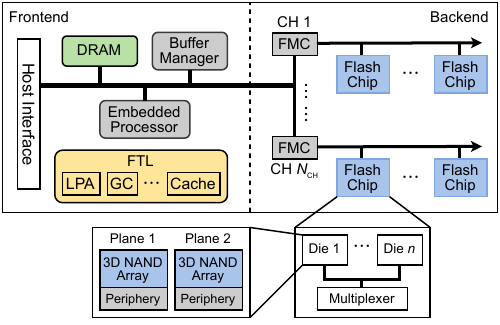}
    \caption{Overall diagram of modern SSD architecture, which is mainly composed of front-end firmware and back-end 3D NAND flash chips.}
    \label{fig:ssd_arch}
\end{figure}

\subsection{Solid-state Drive and Near-Storage Processing}

\noindent\textbf{Modern SSD Architecture.} 
\reffigure\ref{fig:ssd_arch} shows the general architecture of modern SSD that consists of multiple processors, DRAM, and $N_{ch}$ independent I/O channels connecting the controller and flash chips~\cite{chen2011essential,DirikJacob2009}. The frontend has two primary layers called the host interface layer (HIL) and the flash translation layer (FTL). HIL is responsible for managing the host I/O requests~\cite{HIL_1, HIL_2, HIL_3} and FTL handles overall SSD block management including logical-to-physical(L2P) address mapping, garbage collection, and wear leveling. The majority of the internal DRAM capacity is allocated for the logical-to-physical (L2P) mapping table, while the remaining capacity is primarily used for metadata and cache.
Each I/O channel in the SSD backend is equipped with its own flash memory controller (FMC), allowing for operations to be executed across channels independently. Within a channel, multiple flash chips are grouped together to enable parallel read/write operations through a shared bus. This architectural framework facilitates interleaving a wide array of tasks, underpinning its robust functionality and performance capabilities~\cite{SSD_parallelism}. 
The interface is the key to providing high bandwidth for SSD. Throughout this work, we consider NVMe SSDs because NVMe enables significantly faster data transfer rates compared to other interfaces. Specifically, the host interface employs a PCIe interface~\cite{PCIe} and the flash interface employs open NAND flash interfaces (ONFI)~\cite{ONFI4.2}. 


\noindent\textbf{Near Storage Processing.}
Near-storage processing (NSP) architectures are designed to offload computation to the integrated processing engines (PEs) that reside in a storage device. It improves performance for data-intensive applications by mitigating the data movement overhead between storage and processor. NSP is an effective solution to exploit higher internal bandwidth and energy-efficient data path by moving data computation closer to the storage. This approach addresses the ``memory wall~\cite{memorywall}'' problem, caused by limited bandwidth and expensive data movement when running data-intensive workloads. 
The NSP designs vary depending on where the PE is positioned, \ie the PEs can be placed at the FMC level or the flash chip level. The designs at FMC level~\cite{Smart-Infinity_HPCA,SmartSSD,NASCENT,NASCENT2} enjoy the aggregate bandwidth from multiple channels while the designs at flash chip level~\cite{PiF_HotStorage,Parabit,kim2023optimstore,wang2024beacongnn} maximize the bandwidth by aggregating all chip-level bandwidth.

\section{Related Work and Motivation}\label{sec:motivation}

\subsection{Prior Sparse and Compression Algorithms for LLM}
\begin{figure}[ht]
    \centering
    \includegraphics[width=0.7\linewidth]{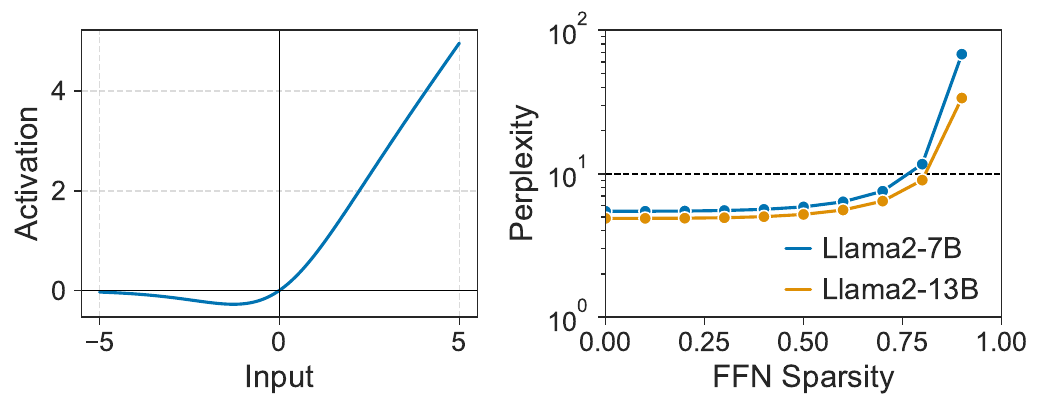}
    \caption{Visualization of SiLU activation (left) and perplexity on WikiText~\cite{wikitext} with various FFN sparsity ratios (right).}
    \label{fig:sparsity_ppl}
\end{figure}

Sparse LLM and model compression are two directions to optimize the performance in environments with limited memory and bandwidth. 
\reftable\ref{tab:comp_spred} summarizes the key properties of state-of-the-art (SoA) zero skipping and compression schemes. 
Recent works~\cite{liu2023dejavu,song2024prosparse,song2023powerinfer,mirzadeh2023relustrikeback} show that LLMs, such as OPT~\cite{zhang2022opt}, with ReLU activation provide opportunities to mitigated redundant computation and weight data movement by masking data access related to zero output from the gate projection. This is realized by adding a sparsity predictor in FFN. 
However, these schemes rely on ReLU activation in FFN to derive sparse activation. ReLU is not available directly in pre-trained SoA LLMs~\cite{touvron2023llama,mixtral,ds_moe} because SiLU~\cite{elfwing2018silu} in \reffigure\ref{fig:sparsity_ppl} is used for better accuracy. 
Although prior works show that SiLU in pre-trained models can replaced with ReLU and model performance can be restored by fine-tuning~\cite{song2024prosparse,mirzadeh2023relustrikeback}, the expensive fine-tuning limits the feasibility and increases the deployment time.

The other direction is to compress the entire LLM model. Matrix decomposition algorithms, such as singular value decomposition (SVD), are used to compress models~\cite{wang2024svd, yuan2023asvd, yu2023compressing, chen2021drone, hsu2207language}. These compression schemes reduce model size and inference computation cost simultaneously, making LLMs easier to fit into hardware with limited memory and computing capability. But the modification to pre-trained weights incurs severe accuracy degradation using high compression ratios. 
The other drawback for mentioned sparse and compression schemes is that they lack the flexibility to trade off between accuracy and latency. This is because the derived sparsity is fixed during training and not runtime configurable.

\subsection{Prior Heterogeneous LLM Accelerators}

\begin{figure}[t]
    \centering
    \includegraphics[width=\linewidth]{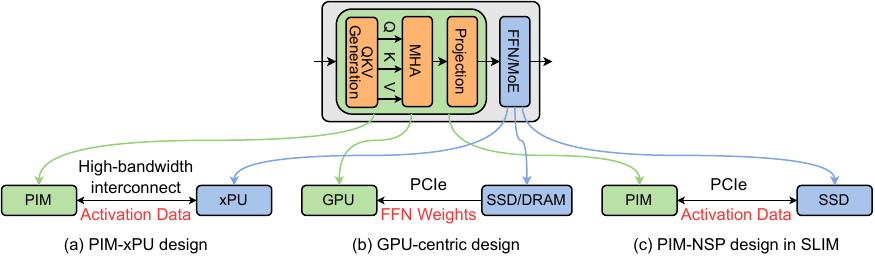}
    \caption{Diagrams of mapping LLM to SoA heterogeneous accelerators, including (a) PIM-xPU and (b) GPU-centric, and (c) \Design's PIM-NSP design for LLM inference. The MHA and FFN/MoE layers in Transformer decoders are heterogeneously mapped to different hardware modules.}
    \label{fig:soa_heter_diagram}
\end{figure}

SoA LLMs~\cite{zhang2022opt,ds_moe,mixtral,touvron2023llama} require billion-scale model sizes to store knowledge. 
The previous work~\cite{park2024attacc} demonstrates that MHA and FFN modules in LLM incur different challenges during inference. MHA is memory-intensive due to the attention and KV cache mechanisms while FFN layers require large memory space to store weights. 
SoA accelerators for LLM inference use heterogeneous computing to offload MHA and FFN modules to different hardware devices, thereby exploiting the advantage of various hardware. 
\reffigure\ref{fig:soa_heter_diagram} summarizes the diagrams of the two popular designs (\textit{PIM-xPU design} and \textit{GPU-centric design}) as follows:
\begin{enumerate}[leftmargin=*]
    \item {\textbf{PIM-xPU Designs.} \reffigure\ref{fig:soa_heter_diagram}-(a) shows the prior PIM-xPU designs~\cite{park2024attacc,heo2024neupims} that take advantage of both PIM and xPU (GPU or NPU) to improve hardware utilization. The rationale is: PIM-enabled HBM has high internal bandwidth which is suitable for accelerating MHA while xPUs are highly optimized for large-batch GEMMs in FFN. Thus, the memory-bound MHA is offloaded to PIM-enabled high-bandwidth memory (HBM) while the computation-intensive FFN is offloaded to xPUs. The intermediate data are transmitted through the high-speed link between PIM and xPU. Existing PIM-xPU designs rely on large batch sizes (64-256) to improve arithmetic intensity and amortize inference cost. In this case, server-grade GPUs and multi-channel PIM modules are needed to provide sufficient memory for both weights and KV cache data. However, the requirements for costly hardware and high service volume are infeasible for local scenarios, making PIM-xPU solutions unsuitable for local deployment.}
    \item \textbf{GPU-centric Solutions.} To reduce the deployment cost, GPU-centric solutions~\cite{sheng2023flexgen,song2023powerinfer,alizadeh2023llminaflash,hwang2023pregate} as shown in \reffigure\ref{fig:soa_heter_diagram}-(b) uses single GPU for inference. Since LLM weights easily exceed single GPU capacity, they are stored in slower but cheaper memories, such as DRAM and SSD. GPU memory only stores KV cache and intermediate data. 
    These designs are significantly more cost-effective compared to PIM-xPU solutions because DRAM and SSD provide sufficient space for storing larger LLMs. 
    However, model weights need to be transferred from SSD/DRAM to GPU memory for inference. The PCIe bandwidth between DRAM/SSD and GPU becomes the bottleneck that limits the system performance. 
\end{enumerate}

\begin{figure}[t]
    \centering
    \includegraphics[width=0.75\linewidth]{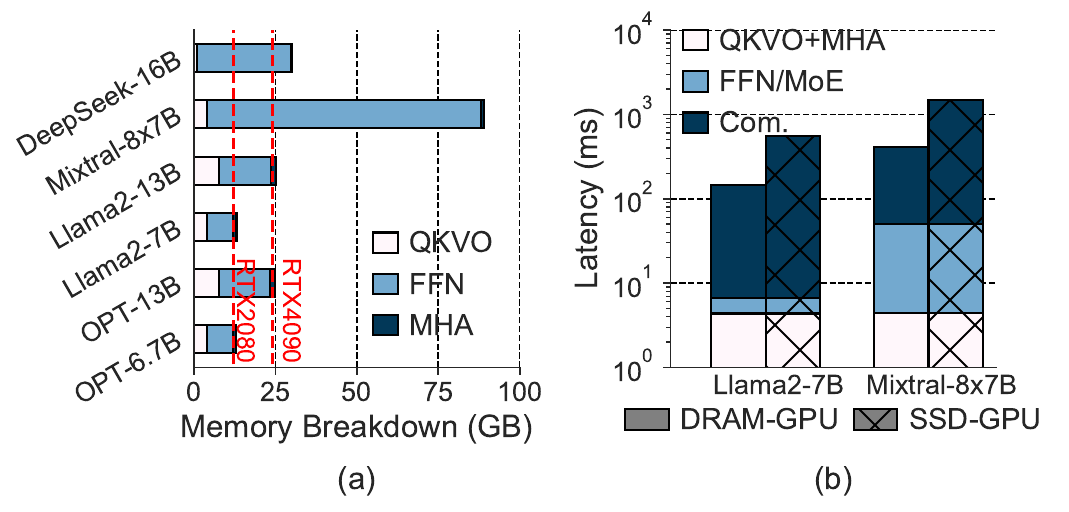}
    \caption{(a) Memory usage breakdown and (b) Runtime breakdown in log scale for LLM inference on GPU-centric systems.}
    \label{fig:profiling_llm_memory_runtime}
\end{figure}

\subsection{Profiling Analysis}
In \reffigure\ref{fig:profiling_llm_memory_runtime}, we perform a profiling analysis to study the memory consumption and runtime statistics for six SoA LLMs.
\reffigure\ref{fig:profiling_llm_memory_runtime}-(a) shows the memory breakdown during inference in FP16 precision. $>60\%$ of memory is consumed by the FFN weights in SoA LLMs. 
Moreover, FFN-related weights in DeepSeek-16B~\cite{ds_moe} and Mixtral-8x7B~\cite{mixtral} dominate the memory usage because of their increasing parameters of MoE layers, \eg $>90\%$ of Mixtral-$8\times7$B parameters coming from FFNs. The substantial memory consumption incur expensive deployment cost because a single commodity GPU is difficult to provide sufficient memory to store the entire model.

GPU-centric solution is regarded as a more cost-effective solution compared to PIM-xPU designs. \reffigure\ref{fig:profiling_llm_memory_runtime}-(b) illustrates the runtime breakdown results (in $\log$ scale) for two GPU-centric designs (DRAM-GPU and SSD-GPU) using the setup in Section~\ref{sec:eval}. The results are sampled from a server with Intel i7-11700K CPU, 64GB DDR4-2400 DRAM with 38.4GB/s bandwidth, and 2TB SSD with PCIe 4.0$\times4$ interface. The used GPU is NVIDIA Geforce RTX 4090 with 24GB memory. The FFN weights are stored in SSD or DRAM while GPU computes QKVO, MHA, and FFN modules. 
It shows that FFN, QKVO, and MHA computation only take a small portion of total runtime. Majority of inference time is consumed by the data communication (Com.) to transfer FFN/MoE weights from DRAM/SSD to GPU memory for both non-MoE (\llama-7B) and MoE model (Mixtral-$8\times7$B). 
There are two main reasons for this: 1. FFN weights consume most of the model size as in \reffigure\ref{fig:profiling_llm_memory_runtime}-(a), 2. the limited PCIe bandwidth significantly increases the data transfer time. Hence, reducing the overhead of FFN or MoE layers is critical to improving inference efficiency.

\subsection{Our Solutions}

\noindent \textbf{Fine-tuning-free Sparsity Prediction.} 
We identify that mitigating FFN data movement is critical for small-batch inference. \Design exploits FFN sparsity to reduce data movement, thus improving inference efficiency. 
We propose a low-rank sparsity predictor with adaptive thresholding in Section~\ref{sec:design_algorithm} to provide low accuracy loss, faster deployment, and better flexibility for runtime tunable sparsity. 


\begin{table}[t]
    \small
    \centering
    \caption{Comparison with existing sparse and compression schemes for LLM.}
    \label{tab:comp_spred}
    \begin{tabular}{c|c|c|c}
         \hline
         \multirow{2}{*}{\textbf{Algorithm}} & MLP-based zero skipping & SVD-based compression & \Design adaptive sparsity prediction \\
         & \cite{song2023powerinfer,song2024prosparse,mirzadeh2023relustrikeback} & \cite{yuan2023asvd,wang2024svd,chen2021drone} & (This work) \\
         \hline
         \textbf{Training (fine-tuning) time} & A few days & 15 mins to 5 hours & $<$10 mins \\
         \hline
         \textbf{Deployment cost} & High & Medium & Low \\
         \hline
         \textbf{Support non-ReLU activation?} & \ding{55} & \ding{51} & \ding{51} \\
         \hline
         \textbf{Runtime sparsity tuning?} & \ding{55} & \ding{55} & \ding{51} \\
         \hline
         \textbf{Accuracy loss} & Low & High & Low \\
         \hline
    \end{tabular}
\end{table}

\noindent \textbf{Offloading to Heterogeneous DRAM-SSD.} 
To address LLM's memory and cost problems, we offload the LLM inference pipeline to the heterogeneous DRAM-SSD system in Section~\ref{sec:design_arch_sys} which takes the advantages of both NSP and PIM. 
As shown in \reffigure\ref{fig:soa_heter_diagram}-(c), the benefits and rationale of proposed \Design include: 1. high-density NAND flashes are mature and low-cost devices to store FFN/MoE weights, since both low-latency or high-density NANDs~\cite{tlcflash,ullflash} can be produced at scale. 2. the NSP-enabled SSD provides both storage and computing capabilities for FFN/MoE layers. The DRAM with PIM capabilities accelerates write-intensive and latency-sensitive MHA. 3. \Design's NSP-PIM heterogeneous system eliminates the bulky weight data transfer over PCIe bus. Only activation need to be synchronized between SSD and DRAM. 


\section{\Design Algorithm for Efficient LLM Deployment}\label{sec:design_algorithm}


\subsection{Low-rank Sparsity Prediction with Thresholding}\label{subsec:lr_sparsity_pred} 
As introduced in Section~\ref{sec:motivation}, existing sparse and LLM compression algorithms~\cite{song2023powerinfer,mirzadeh2023relustrikeback,frantar2023sparsegpt,song2024prosparse} suffer from expensive fine-tuning, limited flexibility or severe accuracy degradation. 
These schemes adopt a Sigmoid function to determine whether the output neurons are predicted to be activated. This lacks runtime flexibility because the sparsity is not tunable. 
The element-wise multiplication in FFN's gating mechanism provides an opportunity to realize more flexible sparsity during runtime: the SiLU output with small absolute values decrease the magnitude of up projection output after element-wise multiplication. We verify this by masking gate output with various thresholds in \reffigure\ref{fig:sparsity_ppl}. LLM performance (in terms of perplexity on WikiText~\cite{wikitext}) gradually degradates when increasing the sparsity ratio of FFN neurons.

\begin{figure}[t]
    \centering
    \includegraphics[width=0.35\linewidth]{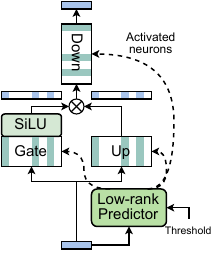}
    \caption{Low-rank FFN sparsity predictor with thresholding.}
    \label{fig:lr_spred}
\end{figure}

\Design has a lightweight sparsity predictor and its generated sparsity is tunable during runtime. 
As depicted in \reffigure~\ref{fig:lr_spred}, a low-rank sparsity predictor is inserted and executed in front of each FFN's gate projection. Meanwhile, the pre-trained weights of LLM model are unchanged to preserve good accuracy. The predictor uses the input sequence $\mathbf{X}$ to predict activated neurons based on the following low-rank approximation with thresholding:
\begin{equation}\label{eq:lr_sparsity_pred}
    \left| \mathbf{X} \cdot \mathbf{L} \cdot \mathbf{R} \right| > threshold,
\end{equation}
where $\mathbf{L}$ is a $dim_e\times dim_{\text{LR}}$ while $\mathbf{R}$ is a $dim_{\text{LR}} \times dim_h$ matrix. 
Instead of hard Sigmoid function, \Design uses a soft threshold value {(as introduced in Section~\ref{subsec:spred_training})} to control the sparsity ratio, thereby achieving tradeoff between accuracy and efficiency. The output of the sparsity predictor are used to bypass inactivated neurons. Only the columns in gate/up weights, and rows in down projection with respect to activated neurons will be fetched and computed.

For each decoder layer, the low-rank matrix $\mathbf{L}$ and $\mathbf{R}$ has a hidden dimension $dim_{\text{LR}}$ which is much smaller than LLM's embedding dimension $dim_e$ (we choose $dim_{\text{LR}} = {dim_e}/{4}$ in this work). 
The additional complexity of sparsity predictor for each layer is $dim_{\text{LR}} \cdot (dim_{e} + dim_{h})$, which is $<10\%$ of original model. Compared to the top-k scheme~\cite{liu2023dejavu} to control sparsity, our scheme is much more hardware friendly since our solution does not require sorter.

\subsection{Training and Threshold Generation} \label{subsec:spred_training}
The training of the proposed sparsity predictor determines the optimal low-rank matrices and sparsity thresholds to realize runtime sparsity configuration. This involves three steps:

\noindent\textbf{Step 1. Calibration data sampling:} 
Previous works~\cite{wang2024svd,yuan2023asvd} show that the FFN input and weights jointly contribute to the dynamic value range of inference, so it is necessary to consider the impact of FFN input. 
To this end, the first step is sampling the calibration data for the input of FFN or MoE layers (denoted as $\mathbf{X}_{\text{sample}}$) that can be used in Step 2. 
\Design reduces the training cost by only sampling calibration data from small-scale dataset, \eg we only take 64 samples from Wikitext~\cite{wikitext}.

\noindent\textbf{Step 2. Activation-aware training:} 
For faster convergence, the low-rank matrices $\mathbf{L}$ and $\mathbf{R}$ are initialized by taking the highest $dim_{\text{LR}}$ ranks from the singular value decomposition (SVD)~\cite{yuan2023asvd,wang2024svd} of FFN gate projection weight $\mathbf{X}_{g}$. 
Then stochastic gradient descent (SGD) and sampled calibration data $\mathbf{X}_{\text{sample}}$ are used to train the sparsity predictor. Similar to~\cite{wang2024svd}, we use the following training loss to minimize the reconstruction error for gate projection module:
\begin{equation}\label{eq:loss}
    \mathcal{L}_{\text{pred}} = \left\lVert \mathbf{X}_{\text{sample}}\cdot \mathbf{W}_g - \mathbf{X}_{\text{sample}}\cdot \mathbf{L}\cdot \mathbf{R} \right\rVert^2.
\end{equation}

\noindent\textbf{Step 3. Generation of sparsity thresholds:} 
After the training is finished, the last step is to determine proper threshold values with respect to each sparsity ratio. We use the sampled calibration data to determine threshold values as $threshold = \text{Top-k}\left( \left| \mathbf{X}_{\text{sample}}\cdot \mathbf{L} \cdot \mathbf{R} \right| \right)$. 
A larger threshold results in higher sparsity at the cost of worse accuracy degradation. The generated threshold values with their corresponding sparsity are stored in a table so that the accelerator can tune the sparsity by using different thresholds during runtime.

\reftable\ref{tab:comp_spred} compares the proposed sparsity predictor to existing sparse predictors~\cite{mirzadeh2023relustrikeback,song2023powerinfer,song2024prosparse} and low-rank compression algorithms~\cite{yuan2023asvd,wang2024svd,chen2021drone}. \Design supports SiLU activation and does not modify the original LLM model. This naturally provides more extensive support for pre-trained LLMs as well as less accuracy loss without model fine-tuning. Meanwhile, \Design speeds up model deployment speed by using a shorter sparsity predictor training time ($<$10mins for \llama-7B). In comparison, other MLP schemes~\cite{mirzadeh2023relustrikeback,song2023powerinfer,song2024prosparse} require to replace SiLU with ReLU, leading to dramatic accuracy loss. They cost days to fine-tune the entire LLM model.

\begin{figure}[t]
    \centering
    \includegraphics[width=0.8\linewidth]{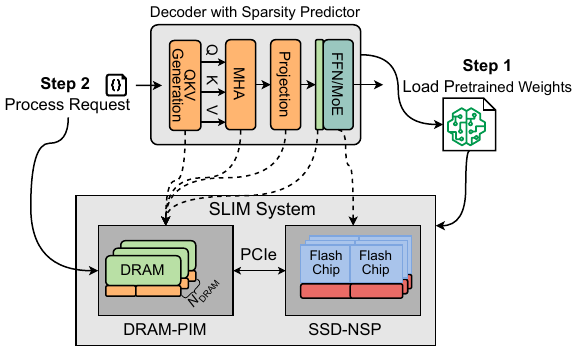}
    \caption{Overview of \Design heterogeneous DRAM-SSD system.}
    \label{fig:overall_design}
\end{figure}

\section{\Designcap Hardware Architecture}\label{sec:design_arch_sys}

In this section, we present \Design, the heterogeneous DRAM-SSD accelerator for efficient LLM serving. 

\subsection{Overview}

\reffigure\ref{fig:overall_design} illustrates the overall diagram of \Design architecture in a heterogeneous DRAM-SSD system that exploits the advantages of DRAM-based PIM and NSP, \ie DRAM-based PIM boosts latency-sensitive sparsity prediction or write-intensive MHA, while NSP handles regular and large-chunk weight data in FFN. 
\Design accelerator is built on the modern SSD and DRAM architecture. A standard NVMe interface connects the DRAM and SSD modules for easy extension and integration. 
To support the whole LLM inference pipeline, \Design system makes three modifications, including: 1. NSP processing engines (PEs) in FMC or NAND die; 2. DRAM-based PIM engine; 3. \Design runtime in FTL firmware.

\noindent\textbf{Heterogeneous Offloading.} 
The first step to offload LLMs in \Design into DRAM-SSD system is to load pretrained weights and sparsity predictor to SSD and DRAM. 
The second step is to send LLM requests to the \Design accelerator for inference. The host sends inference commands and requests data via the NVMe link and host interface. The input sequences are stored in the DRAM before the inference starts. \Design adds several customized commands to SSD's FTL to support these functionalities. To maximize the inference performance, the weight mapping process relies on the optimized weight mapping strategies in Section~\ref{sec:design_software}.

\noindent\textbf{Sparsity Prediction.} 
The latency of sparsity prediction is essential for the inference efficiency of \Design because the generation of addresses and transactions should be fast enough to saturate the back-end NAND bandwidth. The sparsity prediction can be computed using near-storage PEs or DRAM. But near-storage PE execution will experience more complicated FTL layer, leading to higher latency. As in \reffigure\ref{fig:overall_design}, \Design executes sparsity prediction in its DRAM-based PIM engine to ensure low latency and high throughput. The output from sparsity predictor in DRAM-PIM are used as the address to access and control PEs in SSD's NAND chips to only process the activated weights. Other ineffective weights will be skipped.

\subsection{Heterogeneous PIM-NSP Design} 
Since FFN and MoE consume majority of memory space, they are offloaded to the SSD-NSP module that provides high-density storage. On the other hand, QKVO, MHA, and sparsity prediction modules are offloaded to the DRAM-PIM module. 
The computation for MHA differs significantly from FFN in terms of read and write behaviors. 
NSP is suitable for FFN and MoE layers that only require data reading during inference. In comparison, MHA computation involves frequent write and read operations for updating the KV cache data. 
However, writing data to NAND requires 1-3ms erase and 100-600$\mu$s programming time, which dramatically increasing the overall latency. Meanwhile, the frequent and small-chunk KV writing is not suitable for NAND with limited endurance. Meanwhile, the sparsity prediction for controller address and transaction generation is latency sensitive. Therefore, we choose to offload the sparsity prediction, MHA, and QKVO computation into the DRAM-PIM module because DRAM provides low data access and computing latency, which meets the requirements of sparsity prediction and MHA.

\begin{figure}[t]
    \centering
    \includegraphics[width=0.7\linewidth]{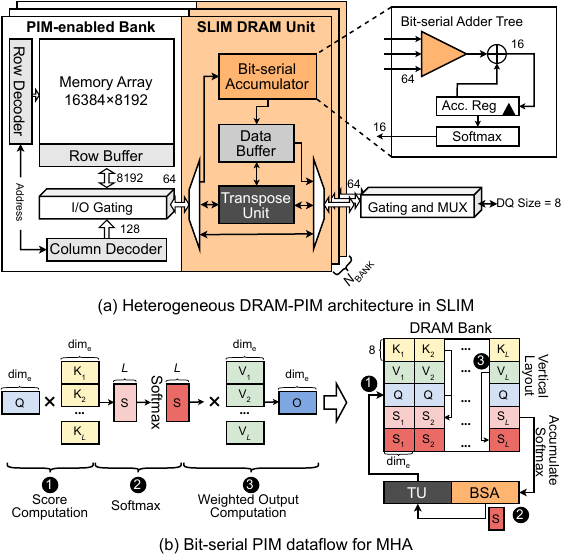}
    \caption{\Design bit-serial DRAM engine.}
    \label{fig:dram_pim}
\end{figure}

\subsection{DRAM-based PIM Engine}\label{subsec:dram_pim}

\noindent\textbf{DRAM-based PIM Engine} leverages PIM to perform in-situ computation for sparsity predictor and MHA. 
\Design replaces the original SSD's DRAM (generally DDR3) with faster DDR4 chips. To save the cost of hardware modifications, \Design's DRAM uses the bit-serial PIM. Previous works~\cite{hajinazar2021simdram,zhou2022transpim} have demonstrated that bit-serial DRAM-based PIM is an energy-efficient and cost-effective solution for GEMM and MHA acceleration. The design details are introduced in Section \ref{subsec:dram_pim}.

\reffigure\ref{fig:dram_pim}-(a) illustrates the architecture of \Design DRAM engine near the SSD controller. There are $N_{\text{DRAM}}$ DDR4-2400 DRAM chips implemented in the engine to provide sufficient capacity to store weights of sparsity predictor, KV cache data, and FTL data. Each DRAM chip utilizes the majority-based bit-serial PIM~\cite{hajinazar2021simdram,ali2019memory,zhou2022transpim} to implement GEMM operations. 
To this end, PIM-related data are stored in bit-serial format, \ie each data point being horizontally organized into consecutive columns. 
The PIM majority operations are performed by issuing two consecutive activate commands to target rows, followed by a precharge command. The PIM majority operations can be consecutively issued to form other complicated arithmetic operations. 
Compared to SoA DRAM designs for LLMs~\cite{park2024attacc,heo2024neupims} which aggressively add bulky GEMM logic gates into DRAM, \Design DRAM engine incurs much less hardware overhead because bit-serial PIM only requires slight changes to DRAM cells and row decoder at the cost of $<1\%$ area overhead~\cite{hajinazar2021simdram,seshadri2017ambit}.

\textbf{Near-bank Unit.} 
Bit-serial PIM's computing efficiency mainly comes from the extensive parallelism over the DRAM column dimension. The high parallelism can amortize the long latency incurred by DRAM commands. However, the main bottleneck of bit-serial-based DRAM is the runtime overhead to create the layout satisfied bit-serial computation~\cite{zhou2022transpim}. \reffigure\ref{fig:dram_pim}-(a) show the architecture of DDR4 DRAM. The slow layout time mainly stems from the limited 8b DQ size, which exposes low external bandwidth. To reduce the latency due to bit-serial data layout. \Design implements the near-bank unit to extend the internal DRAM bandwidth for data layout process. A data buffer and transpose unit in each bank are used to realize fast data layout. The internal bandwidth equals to $64\times N_{\text{bank}}$b per cycle, achieving $8\times N_{\text{bank}}$ speedup.

\textbf{MHA Computation.} 
\reffigure\ref{fig:dram_pim}-(b) shows how MHA is computed in \Design DRAM engine. The TU first copies and converts the newly generated query $Q$ vector to bit-serial format, \ie copies $Q$ vector by $L$ times and align to the rows. \circled{1}-Attention score computation step issues DRAM commands to perform bit-serial pointwise multiplications.  
\circled{2}-Softmax calculation.
After the element-wise multiplication is finished, BSA reads the results and then calculates the accumulation and Softmax values. 
In \circled{3}-Output calculation step, the attention score $S$ is sent back to TU and perform the second bit-serial in memory. Finally, the weighted sum of normalized score $S$ on $V$ is computed.

\begin{figure}[t]
    \centering
    \includegraphics[width=0.75\linewidth]{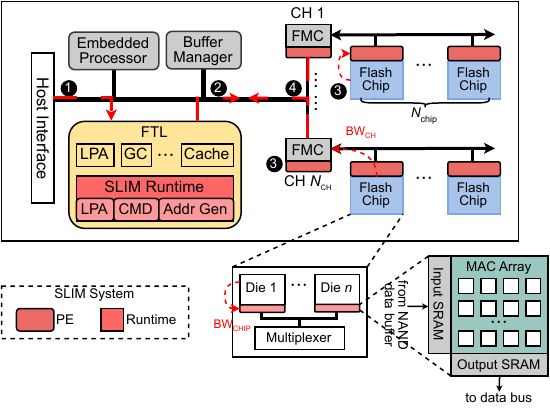}
    \caption{SSD with chip-level NSP capabilities in \Design.}
    \label{fig:ssd_nsp}
\end{figure}

\subsection{SSD with NSP Capabilities}\label{subsec:nsp_pe}
After receiving the commands and request data, \Design FTL runtime schedules and manages the LLM inference execution in four steps as depicted in \reffigure\ref{fig:ssd_nsp}. 
\circled{1}-Data broadcast step copies and sends $B\times L\times dim_e$ input sequences to all input SRAM buffers of $N_{\text{FMC}}$ FMCs or $N_{\text{FMC}}\times N_{\text{chip}}$ NAND flash chips. 
\circled{2}-Address and transaction information is generated to translated logical address to physical space so that FFN weights can be fetched from NAND. \Design FTL runtime is in charge of LPA calculation and associated transaction generation. The generated transactions are delivered to the transaction scheduling unit (TSU) for the following processings.
\circled{3}-NSP GEMM computation step issues commands to PEs in FMC or die to compute FFN, and MoE. PEs at the FMC or die level receive the fetched weights and perform GEMM operations. The resulting partial sum results are cached in each PE's output SRAM buffer. 
\circled{4}-Reduce and collect step is responsible for collecting and accumulating the partial sum results in the output SRAM buffers of different PEs. This is achieved by issuing the reduce commands to PEs. Each PE sends back the corresponding partial sum via channel bus (die-level PE) or on-chip bus (channel-level PE).

\noindent\textbf{Near-storage Processing Engine (PE)} is used to compute GEMMs near LLM weight storage. We implement two ASIC PE designs at different SSD memory hierarchies, \ie channel level or die level. These two designs trade off hardware overhead and achievable memory bandwidth. Channel and die-level designs are used to perform near-storage GEMMs for FFN. As shown in \reffigure\ref{fig:ssd_nsp}, PEs are equipped with two SRAM buffers and a MAC array. 
\Design PEs adopt the output stationary dataflow, and the input SRAM caches all $B\times L\times dim_e$ input sequences with batch size $B$ and sequence length $L=1$. In this case, the corresponding weights only need to be fetched once from NAND, thus avoiding repeated data movement. 

\noindent\textbf{Channel-level Design} places PEs in FMC modules to exploit the aggregate bandwidth of different channels. Each channel-level PE is only able to process the data coming from NAND chips in the associated channel. In this case, the aggregate peak bandwidth is $N_{\text{CH}}\times BW_{\text{CH}}$, where $BW_{\text{CH}}$ is determined by the channel I/O bus used, such as ONFI~\cite{ONFI3.1,ONFI4.2}. 

\noindent\textbf{Die-level Design} implements PEs in each NAND die as illustrated in \reffigure\ref{fig:ssd_nsp}. Die-level PE multiplexes the data bus between NAND data register and ONFI bus. The aggregate peak bandwidth exposed to die-level design is $N_{\text{CH}}\times N_{\text{chip}} \times BW_{\text{DIE}}$, where $BW_{\text{DIE}}$ and $N_{\text{chip}}$ denote the NAND I/O rate and the number of chips under a channel, respectively. This suggests that die-level design usually has a higher internal bandwidth compared to channel-level design. But each die-level PE can only access the data on its dies.

\begin{figure}[t]
    \centering
    \includegraphics[width=0.7\linewidth]{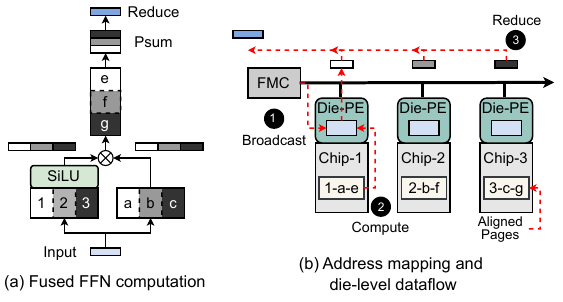}
    \caption{Fused FFN computation and page-aligned address mapping for LLM weights.}
    \label{fig:fused_data_mapping}
\end{figure}

\noindent\textbf{Fused FFN Computation.} 
LLM sparsity allows us to skip sparse column indices in gate matrix $\mathbf{W}_g$ and up projection matrix $\mathbf{W}_u$. Additionally, the corresponding row in down projection matrix $\mathbf{W}_d$ can also be skipped. To this end, \Design PEs adopt the fused FFN computation to fully leverage LLM sparsity. 
The computation of PEs in \reffigure\ref{fig:fused_data_mapping}-(a) does not follow the order of gate, up projection, and down projection weight matrices. Instead, the column vector in gate and up matrices together with the row vectors in down projection matrix have highest priority. Specifically, the PE first consecutively processes vectors 1-a-e to produce the first partial sum (Psum). Next, the 2nd columns of $\mathbf{W}_g$ and $\mathbf{W}_u$ together with 2nd row of $\mathbf{W}_d$ are computed. In this case, PE do not need to store intermediate results with a high dimension $dim_h$. The other benefit is that it increases the request size, allowing each page to encompass more effective data.

\begin{figure}[t]
    \centering
    \includegraphics[width=0.6\linewidth]{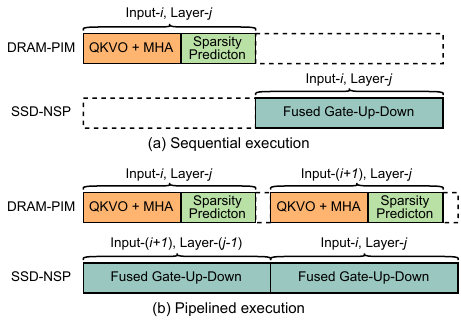}
    \caption{Sequential and pipelined execution flows in \Design.}
    \label{fig:pip_flow}
\end{figure}

\section{\Designcap Optimization and Integration}\label{sec:design_software}

\subsection{Weight Mapping Scheme}\label{subsec:opt_data_mapping}

\noindent\textbf{Page-aligned Mapping.}
The scheme to map pretrained LLM weights onto SSD is critical to maximize \Design performance. 
For SSD, ensuring sequential and aligned reading of weight data is pivotal to maximize read bandwidth because misaligned data can result in increased read requests. The other consideration is the limitations on the accessible memory space of different PE designs, \ie die-level PE can only access the data on its die. To address this challenge, we modify the generic address mapping that interleaves consecutive pages across channels. 
\Design FTL in \reffigure\ref{fig:fused_data_mapping}-(b) distributes each fused vector to different dies in a page-aligned manner. For example, 3-c-g is stored in single or consecutive pages in a die. 
This ensures that all weights are aligned and evenly distributed in different dies, which is helpful to maximize parallelism of NSP.

\noindent\textbf{Vector Packing.}
We also observe that most SSD page sizes ranging from 4KB to 16KB~\cite{ullflash,tlcflash} are equal to or larger than the weight vector size $dim_e$. 
For large page size, such as 16KB, the fused vector with small $dim_e$ is unable to fully fill the allocated pages, causing both performance degradation and storage underutilization. 
This is because each page read takes at least $t_R$ execution time. We improve page utilization by packing multiple continuous fused vectors when $dim_e$ is less than page size.

\subsection{Pipelined Execution}\label{subsec:opt_pipeline}
\reffigure\ref{fig:pip_flow} compares the timing diagram for sequential and pipelined executions. Without any optimizations, the DRAM-PIM and SSD modules are executed in a sequential manner in \reffigure\ref{fig:pip_flow}-(a) due to the data dependency between MHA and FFN modules. The sequential execution time is estimated as $t_{\text{DRAM}} + t_{\text{SSD}}$ and the hardware utilization is low. 
The performance and utilization can be improved by simultaneously executing DRAM-PIM and SSD modules in a pipelining manner as shown in \reffigure\ref{fig:pip_flow}-(b). Specifically, the DRAM-PIM and SSD-NSP executions for two independent input sequences can be interleaved. The resulting average throughput is improved by a factor of $\dfrac{t_{\text{DRAM}} + t_{\text{SSD}}}{\max(t_{\text{DRAM}}, t_{\text{SSD}})}$.

\subsection{FTL Modifications}
\Design requires a modification of the conventional FTL to introduce a specialized inference mode tailored for LLM tasks and manage address space. Since the read/write operation of SSD through NVMe protocol is also necessary for general purposes or updating the models, supporting custom commands without affecting the read/write commands is important. To effectively manage both the generic and custom commands, \Design runtime manages additional LPA-to-buffer mapping information as illustrated in \reffigure\ref{fig:ssd_nsp}, similar to the idea of the translation lookaside buffer. The memory complexity to store LPA entries for the proposed weight mapping scheme is $\mathcal{O}(N_{\text{dec}} \cdot dim_h)$ for FFN, where the number of decoder layers $N_{\text{dec}}<100$ while the hidden dimension $dim_h$ ranges from 1K to 14K. The required complexity is negligible as compared to the generic LPA. 
In this manner, \Design can operate in both modes without redundant reads from flash chips and duplicating the L2P mapping. We also consider the delays in other execution while the internal DRAM bandwidth is consumed by computation, preventing the conflict of device resources.



\begin{table}[t]
    \small
    \centering
    \caption{Hardware specifications of \Design.}
    \renewcommand{\arraystretch}{1.1}
        \begin{tabular}{ c|p{6cm}  }
            \hline
            \multicolumn{2}{c}{\textbf{SSD and NAND Parameters}} \\
            \hline
            \textbf{Host Interface} & NVMe, PCIe 4.0$\times4$ with 8GB/s bandwidth\\
            \hline
            \textbf{NAND I/O} & NV-DDR3 1200MT/s, 8b channel bus \\
            \hline
            \multirow{5}{*}{\textbf{Low-latency SSD}~\cite{ullflash}} & 240GB SLC \\
            & 16 channels, 4 chips, 1 dies, 2 planes\\
            & 1024 blocks, 512 pages, 4KB page size \\
            & \textbf{Timing:} $t_{R}$ = 3$\mu$s, $t_{PROG}$ = 100$\mu$s\\
            & \textbf{Area:} 50.63 mm$^2$/die\\
            \hline
            \multirow{5}{*}{\textbf{High-density SSD}~\cite{tlcflash}} & 1TB TLC \\
            & 16 channels, 4 chips, 1 dies, 2 planes\\
            & 1024 blocks, 512 pages, 16KB page size \\
            & \textbf{Timing:} $t_{R}$ = 40$\mu$s, $t_{PROG}$ = 650$\mu$s\\
            & \textbf{Area:} 15.06 mm$^2$/die\\
            \hline
            \hline
            \multicolumn{2}{c}{\textbf{DRAM Parameters}}\\
            \hline
            \multirow{3}{*}{\textbf{Configurations}} & 32GB DDR4-2400 DRAM @ 1.2GHz\\
            & 32 chips, 4 bank groups, 4 banks/bank group \\
            & 65536 rows, 1024 columns, 8KB page size\\
            \hline
            \multirow{3}{*}{\textbf{Timing Parameters}} & nRCD = 18, nRAS = 39, nRP = 18, \\
            & nCCD\_S=4, nCCD\_L=6, nFAW=40, \\
            &  nCL = 18, nWR = 18, nCCD = 4\\
            \hline
            \hline
            \multicolumn{2}{c}{\textbf{ASIC Parameters}}\\
            \hline
            \multirow{3}{*}{\textbf{Channel-level PE}} & \textbf{Config:} 64 MACs, 128KB SRAM \\
            & \textbf{Area:} 0.15 mm$^2$/channel \\
            & \textbf{Power}: 60.7 mW/channel @ 1GHz\\
            \hline
            \multirow{3}{*}{\textbf{Die-level PE}} & \textbf{Config:} 16 MACs, 64KB SRAM \\
            & \textbf{Area:} 0.07 mm$^2$/die \\
            & \textbf{Power}: 27.0 mW/die @ 1GHz\\
            \hline
            \multirow{3}{*}{\textbf{Near-DRAM Unit}} & \textbf{Config:} 256KB SRAM, 64b bit-serial adder tree \\
            & \textbf{Area:} 0.28mm$^2$/bank \\
            & \textbf{Power}: 32.3mW/bank @ 1GHz\\
            \hline
        \end{tabular}
    \label{tab:hardware_specs}
\end{table}

\section{Evaluation}\label{sec:eval}
\subsection{Methodology}

\noindent\textbf{System Setup.}
The evaluation system is equipped with Intel i7-11700K and 2-channel 64GB DDR4-2400 DRAM that provides 38.4GB/s bandwidth. NVIDIA Geforce RTX 4090 with 24GB memory and 1TB/s bandwidth is used. We measure the power consumption of the GPU using \texttt{nvidia-smi} and use the data provided in~\cite{wang2024beacongnn} to measure the energy of the PCIe link.

\noindent\textbf{Baselines.} 
We consider two GPU-centric designs as follows:
\begin{itemize}[leftmargin=*]
    \item \textbf{SSD-GPU:} Similar to~\cite{alizadeh2023llminaflash}, weights of FFN or MoE layers are stored in SSD and transferred to GPU via PCIe 4.0$\times4$ link. SSD uses the high-end parameters given in \reftable\ref{tab:hardware_specs}.
    \item \textbf{DRAM-GPU:} Similar to~\cite{song2023powerinfer}, weights related to FFN or MoE are stored in 64GB host DDR4-2400 DRAM and transferred to GPU via PCIe 4.0$\times16$ link. 
\end{itemize}
The sparsity prediction, QKVO, and MHA are computed by GPU using Pytorch~\cite{pytorch}.

\noindent\textbf{SSD and Hardware Modeling.}
We adopt two distinct types of NVMe SSDs for our comparative analysis: a low-latency SSD employing SLC technology~\cite{ullflash} and a high-density SSD utilizing TLC technology~\cite{tlcflash}, as detailed in Table \ref{tab:hardware_specs}. The connection to the host is facilitated through a PCIe 4.0 interface by 4 lanes and the connection to the flash chip is through 8 channels of ONFI NVDDR3 with a data transfer rate of 1200MT/s.

The ASIC hardware is implemented using Verilog HDL and synthesized by Synopsys Design Compiler using TSMC 40nm technology. The clock frequency is 1GHz and the design is scaled to 22nm using methods in~\cite{sarangi2021deepscaletool}. Timing and energy parameters of SRAM buffers are calculated using CACTI-3DD~\cite{CATCI-3DD} in 22 nm.

\noindent\textbf{Simulation Methodology.}
The performance \Design is evaluated using a modified simulator from MQSim~\cite{tavakkol2018mqsim} and TransPIM~\cite{zhou2022transpim}. The simulator realizes event-driven and cycle-accurate simulation for SSD-NSP and DRAM-PIM modules. We add \Design commands to FTL and update the corresponding device timing using parameters in \reftable\ref{tab:hardware_specs}. The simulation traces are collected from each evaluation dataset. The energy numbers for NAND and PCIe are from previous work~\cite{wang2024beacongnn}.

\begin{table}[t]
    \small
    \centering
    \caption{Specifications of evaluated LLMs.}
    \renewcommand{\arraystretch}{1.1}
        \begin{tabular}{c|c|c|c|c}
            \hline
            \textbf{Type} & \textbf{Model} & \textbf{\# of Layer} & $\mathbf{dim}_e$ & \textbf{Activated/Total FFNs per MoE} \\
            \hline
            \multirow{2}{*}{FFN}& Llama-2-7B~\cite{touvron2023llama} & 32 & 4096 & - \\
            & Llama-2-13B~\cite{touvron2023llama} & 40 & 5120 & - \\
            \hline
            \multirow{2}{*}{MoE}& DeepSeek-16B~\cite{ds_moe} & 24 & 2048 & 8/64 \\
            & Mixtral-8$\times$7B~\cite{mixtral} & 32 & 4096 & 2/8 \\
            \hline
        \end{tabular}
    \label{tab:eval_llm_spec}
\end{table}

\noindent\textbf{Benchmarks.} 
We evaluate various LLMs with parameters from 7B up to 56B. \reftable\ref{tab:eval_llm_spec} shows the model specifications of evaluated LLMs. All models in our experiments use 8b quantization. We evaluate the zero-shot accuracy for LLMs using the following benchmarks: commonsense reasoning  (WinoGrande~\cite{sakaguchi2019winogrande}, HellaSwag~\cite{zellers2019hellaswag}, PIQA~\cite{bisk2020piqa}, BoolQ~\cite{BOOLQ}, and OpenBookQA~\cite{OpenBookQA2018}) and reading comprehension (Arc-Easy and Arc-Challenge~\cite{Clark2018ThinkYH}). Throughout the evaluation, we use a sequence length of $L=2048$ by default.

\subsection{Accuracy Evaluation}

\begin{table}[t]
    \small
    \centering
    \caption{Zero-shot accuracy of \llama~\cite{touvron2023llama} with \Design sparsity predictor. Various sparsity ratios are used. {$0\%$ is the dense baseline.}}
    \begin{tabular}{c|c||c|c|c|c|c|c|c|c}
        \hline
        \textbf{Model} & \textbf{Sparsity Ratio} & \textbf{Arc-C} & \textbf{Arc-E} & \textbf{BoolQ} & \textbf{Hswag} & \textbf{OBQA} & \textbf{PIQA} & \textbf{Wino.} & \textbf{Avg.} \\
        \hline
        \multirow{4}{*}{\llama-7B} 
        &  $0\%$ & 0.45 & 0.71 & 0.76 & 0.68 & 0.50 & 0.80 & 0.68 & 0.65\\
        & $20\%$ & 0.45 & 0.70 & 0.74 & 0.68 & 0.50 & 0.80 & 0.68 & 0.65 \\
        & $40\%$ & 0.45 & 0.69 & 0.73 & 0.68 & 0.48 & 0.79 & 0.65 & 0.64 \\
        & $60\%$ & 0.44 & 0.62 & 0.70 & 0.66 & 0.47 & 0.80 & 0.63 & 0.62 \\
        \hline
        \multirow{4}{*}{\llama-13B} 
        &  $0\%$ & 0.46 & 0.75 & 0.83 & 0.70 & 0.49 & 0.81 & 0.72 & 0.67 \\
        & $20\%$ & 0.44 & 0.74 & 0.82 & 0.71 & 0.47 & 0.81 & 0.72 & 0.67 \\
        & $40\%$ & 0.44 & 0.74 & 0.79 & 0.70 & 0.48 & 0.80 & 0.68 & 0.66 \\
        & $60\%$ & 0.42 & 0.69 & 0.79 & 0.69 & 0.46 & 0.79 & 0.68 & 0.65 \\
        \hline
    \end{tabular}
    \label{tab:eval_llm_sparsity}
\end{table}

\noindent \textbf{Accuracy with Different Sparsity Ratios.} 
We evaluate the zero-shot accuracy of \llama-7B and 13B models across various tasks, as presented in \reftable\ref{tab:eval_llm_sparsity}. Different threshold values are chosen to derive different sparsity ratios from 0$\%$ to 60$\%$. Negligible performance loss is observed when using a sparsity ratio $<60\%$. Increasing sparsity ratio to $60\%$ leads to 2-3$\%$ average zero-shot accuracy loss, which is less than other SVD-based algorithms~\cite{yuan2023asvd,wang2024svd}. Meanwhile, we can set different threshold values to generate different sparsity levels, thus trading off between inference efficiency and accuracy.

\begin{table}[t]
    \small
    \centering
    \caption{{Accuracy loss comparison of \llama-7B~\cite{touvron2023llama} for sparse LLMs based on low-rank decomposition.}}
    \begin{tabular}{c|c||c|c|c|c|c}
        \hline
        \textbf{Sparsity Ratio} & \textbf{Algorithm} & \textbf{Hswag} & \textbf{OBQA} & \textbf{PIQA} & \textbf{Wino.} & \textbf{Avg.} \\
        \hline
        \multirow{4}{*}{$20\%$}
        & FWSVD~\cite{hsu2207language} & 0.49 & 0.25 & 0.69 & 0.65 & 0.52 \\
        & ASVD~\cite{yuan2023asvd}     & 0.16 & 0.05 & 0.11 & 0.06 & 0.10 \\
        & SVD-LLM~\cite{wang2024svd}   & 0.02 & 0.01 & 0.00 & 0.01 & 0.01 \\
        & \Design (This Work)          & 0.00 & 0.00 & 0.00 & 0.00 & 0.00 \\
        \hline
        \multirow{4}{*}{$40\%$}
        & FWSVD~\cite{hsu2207language} & 0.57 & 0.28 & 0.74 & 0.68 & 0.57 \\
        & ASVD~\cite{yuan2023asvd}     & 0.49 & 0.26 & 0.66 & 0.61 & 0.51 \\
        & SVD-LLM~\cite{wang2024svd}   & 0.05 & 0.05 & 0.10 & 0.02 & 0.06 \\
        & \Design (This Work)          & 0.00 & 0.02 & 0.01 & 0.03 & 0.02 \\
        \hline
        \multirow{4}{*}{$60\%$}
        & FWSVD~\cite{hsu2207language} & 0.56 & 0.28 & 0.77 & 0.69 & 0.58 \\
        & ASVD~\cite{yuan2023asvd}     & 0.48 & 0.29 & 0.71 & 0.64 & 0.53 \\
        & SVD-LLM~\cite{wang2024svd}   & 0.26 & 0.16 & 0.44 & 0.26 & 0.28 \\
        & \Design (This Work)          & 0.02 & 0.03 & 0.00 & 0.05 & 0.03 \\
        \hline
    \end{tabular}
    \label{tab:eval_llm_acc_loss}
\end{table}

\noindent \textbf{Accuracy Loss Comparison with Other Low-rank Methods.} 
Table~\ref{tab:eval_llm_acc_loss} compares the accuracy degradation to various low-rank-based sparse LLM algorithms. Our proposed \Design demonstrates superior accuracy preservation compared to other approaches, including FWSVD~\cite{hsu2207language}, ASVD~\cite{yuan2023asvd}, and SVD-LLM~\cite{wang2024svd}. At the 20\% sparsity threshold, \Design achieves zero accuracy loss across all evaluated benchmarks, while FWSVD exhibits substantial degradation with an average accuracy loss of 0.52. This performance advantage persists across increasing sparsity ratios, with \Design maintaining minimal degradation of 0.03 average accuracy loss at 60\% sparsity. The main difference between \Design and other SVD baselines is that \Design does not compress FFN weights, thereby reducing accuracy loss. In comparison, \Design achieves sparse LLM inference by skipping those predicted neurons with small or zero values. The consistent superiority of \Design across different datasets demonstrates its efficacy in achieving aggressive sparsity while preserving inference accuracy.

\begin{figure}[t]
    \centering
    \includegraphics[width=0.35\linewidth]{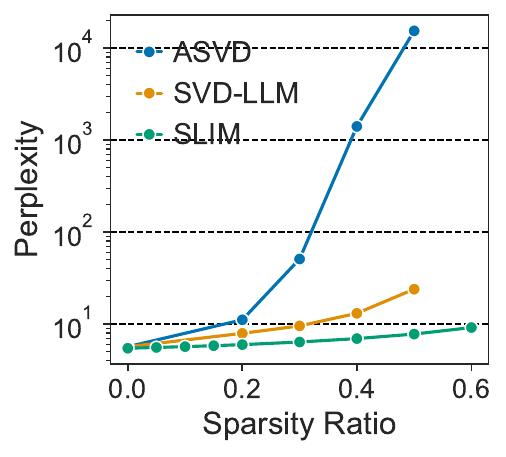}
    \caption{Perplexity performance of various algorithms when using different sparsity ratios.}
    \label{fig:perf_ppl}
\end{figure}

\noindent \textbf{Perplexity Performance.} 
We compare \Design's sparsity prediction scheme to other SVD-based compression methods, ASVD~\cite{yuan2023asvd} and SVD-LLM~\cite{wang2024svd}. 
Perplexity on WikiText~\cite{wikitext} is adopted as the performance metric. \reffigure\ref{fig:perf_ppl} shows that \Design consistently achieves lower perplexity than ASVD and SVD-LLM. This indicates that \Design suffers from less capability degradation in terms of autoregressive modeling.     
It should be noted that the FFN sparsity in ASVD and SVD-LLM are controlled by setting different SVD hidden dimensions during training. In comparison, \Design is more flexible since the sparsity can be tuned by setting different runtime thresholds for the sparsity predictor.

\subsection{Performance Evaluation}

\begin{figure}[t]
    \centering
    \includegraphics[width=0.75\linewidth]{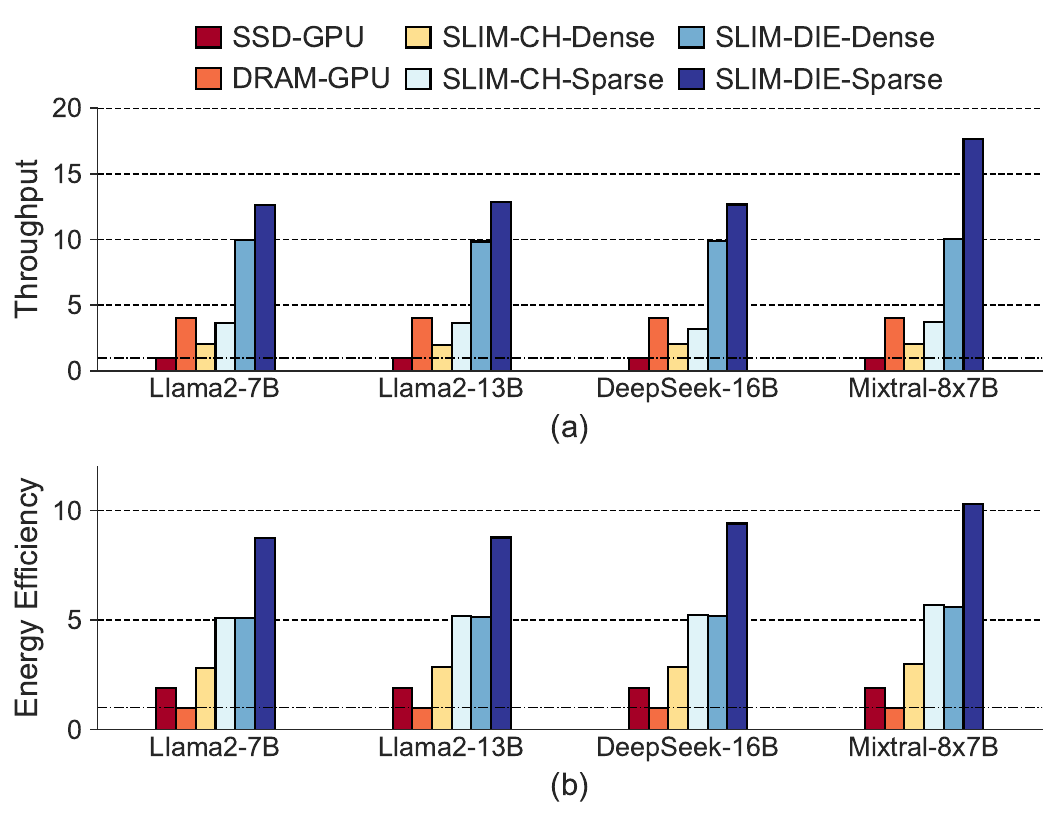}
    \caption{(a) Normalized throughput and (b) normalized energy efficiency comparison for SSD-GPU, DRAM-GPU, \Design-CH, and \Design-DIE on SoA LLMs. Sparsity ratio is set to 0.5.}
    \label{fig:norm_tp_energy}
\end{figure}

\noindent\textbf{Throughput.}
We compare the throughput of \Design channel-level and die-level designs with SSD-GPU and DRAM-GPU baselines for various SoA LLMs including Llama-2, DeepSeek and Mixtral. In this experiment, \Design uses SLC as the NAND device. 
\reffigure\ref{fig:norm_tp_energy}-(a) summarizes the normalized throughput comparison on the LLMs. \Design die-level design essentially achieves the highest throughput among all the hardware configurations. With the employment of sparsity prediction, the performance of \Design die-level designs are further improved by $1.28\times$ to $1.76\times$. Although \Design channel-level designs leverage PIM and NSP. the throughput is limited by the channel-level ONFI bus bandwidth, thus provides worse performance than DRAM-GPU baseline. Still, it generally outperforms SSD-GPU as a low-cost alternative. As mentioned in Section~\ref{sec:eval}, SSD-GPU suffers from limited bandwidth of SSD compared to DRAM, thus providing the worst throughput. Overall, \Design die-level design with sparsity achieves $17.6\times$ and $4.3\times$ comparing to SSD-GPU and DRAM-GPU baslines.

\noindent\textbf{Energy Efficiency.}
We calculate the normalized energy efficiency of \Design and the mentioned baselines for LLM generation in \reffigure\ref{fig:norm_tp_energy}-(b). \Design designs have a better energy efficiency on all tested LLMs. This is because the NSP and PIM capabilities of \Design reduces the energy consumption required to perform data movement through in-situ computation or near-storage processing. 
Furthermore, \Design die-level design with sparsity prediction enabled shows a superior energy efficiency, \ie $8.8\times$ to $10.3\times$ better energy efficiency over the DRAM-GPU. The benefits are attributed to the reduced FFN/MoE weight fetching thanks to the sparsity prediction mechanism. We notice that the energy efficiency improvements on MoE-based models (DeepSeek-16B and Mixtral-8$\times$7B) are slightly better than non-MoE models. This is because MoE-based models consume more energy on moving and processing MoE weights than non-MoE models. The energy reduction due to sparsity is more significant.

\begin{figure}[t]
    \centering
    \includegraphics[width=0.75\linewidth]{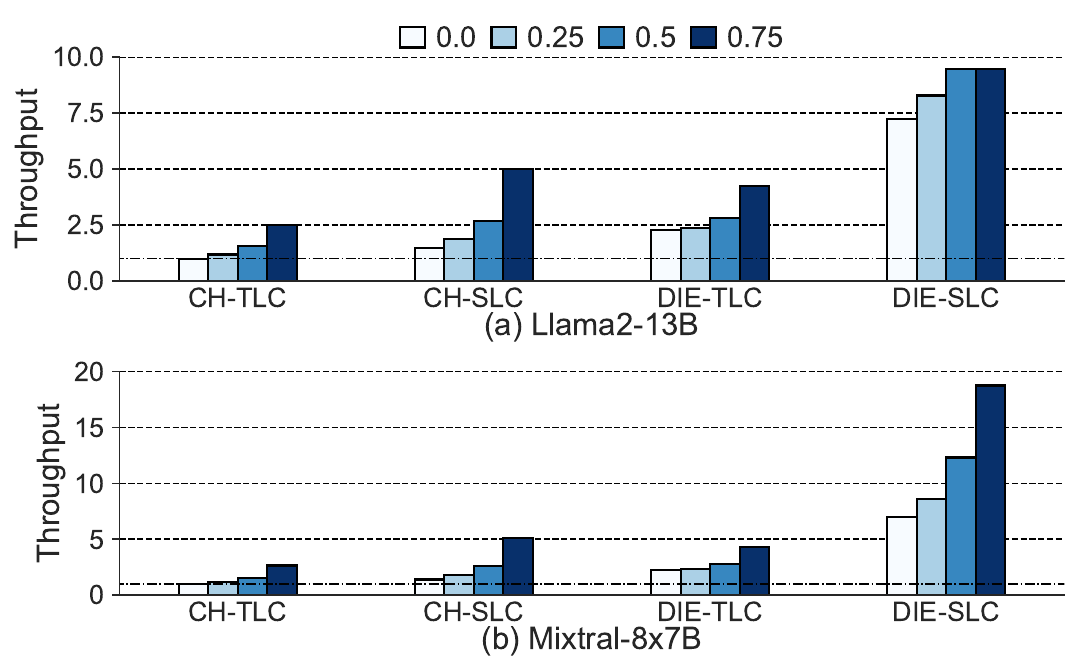}
    \caption{Performance comparison of \Design channel-level (CH) and die-level (DIE) designs with low-latency SLC or high-density TLC NAND devices. The sparsity ratios vary from 0 to 0.75.}
    \label{fig:perf_ch_die_comp}
\end{figure}

\subsection{Sensitivity Analysis}

In this section, we study various design choices and factors that impact the overall inference performance, such as designs levels, NAND devices, and sparsity ratios.

\noindent \textbf{Different Designs Levels and NAND Devices.} 
Our proposed channel-level and die-level designs enjoy different aggregate bandwidths from channels or chips. We evaluate and compare the performance difference of these two designs on models \llama-13B and Mixtral-8$\times$7B using various sparsity ratios. \reffigure\ref{fig:perf_ch_die_comp} shows the comparison of normalized throughput for \Design channel-level and die-level designs with SLC and TLC NAND flash chips. The comparison shows that the low-latency SSD outperforms high-density SSD in both channel-level and die-level design. 

When comparing different design levels that use the same NAND device, \Design die-level design has 1.6 to 5$\times$ throughput improvements over the channel-level design since the die-level design achieves a higher aggregate bandwidth compared to the channel-level design. 
But the SLC device shows more significant speedup over TLC when switching from channel level to die channel. Using the SLC device, the die-level design show $>4\times$ speedup over the channel-level design. In comparison, the die-level design only generates 1.6 to 2.3$\times$ speedup when using the TLC device.  This is because the peak read bandwidth of TLC flash chip is only $\approx\frac{1}{3}$ of the SLC counterpart. 

The tradeoff between storage density and inference performance can be achieved by choosing high-density TLC or low-latency SLC NAND device. For the channel-level design, the throughput improvement of using SLC is only 1.4 to 2.0$\times$, which is lower than the relative bandwidth ratio ($\approx3\times$) between SLC and TLC devices. This is because the channel-level design is mainly bounded by the FMC I/O rate ($\approx1.2$GB/s) for each channel. 

\noindent\textbf{Impact of Sparsity.}
\reffigure\ref{fig:perf_ch_die_comp} also shows the normalized throughput comparisons for \Design using various sparsity ratios. The throughput increases with higher sparsity for both channel-level and die-level designs. 
Compared to the channel-level design, the die-level design fully exploits the advantages of SLC device. The DIE-SLC design shows $\approx3.2\times$ speedup for dense inference and up to $5.0\times$ speedup for sparse inference over DIE-TLC. The SLC device has lower page read latency and smaller page size over TLC. So it is more suitable for sparse LLM inference, where random and small weight vector readout is more frequent.

\begin{figure}[t]
    \centering
    \includegraphics[width=0.75\linewidth]{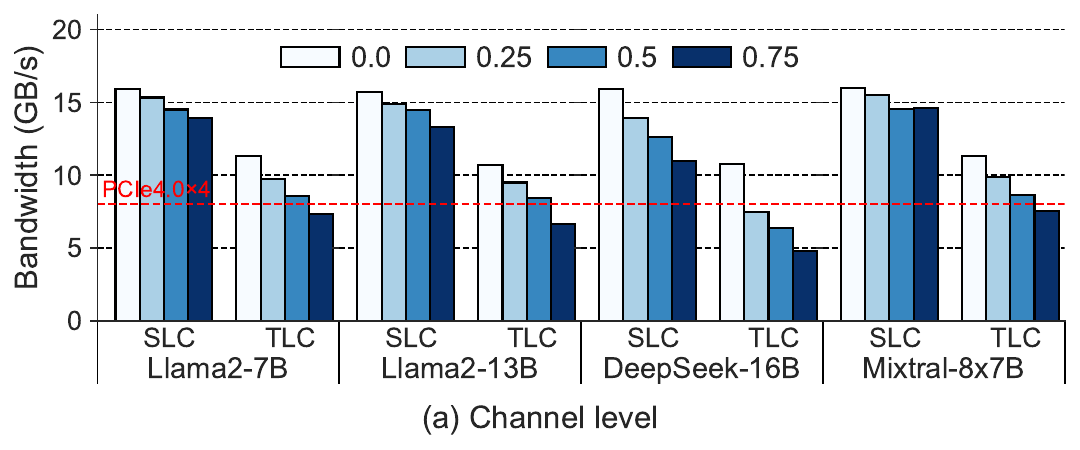}
    \includegraphics[width=0.75\linewidth]{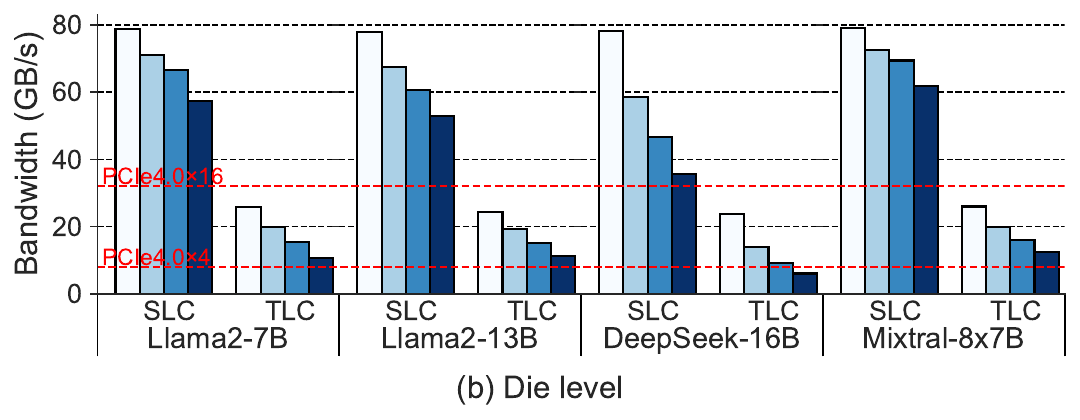}
    \caption{Comparison for the aggregate read bandwidth of \Design channel-level (CH) and die-level designs with low-latency or high-density NAND devices. The sparsity ratios vary from 0 to 0.75.}
    \label{fig:perf_tlc_slc_comp}
\end{figure}

\noindent\textbf{Read Bandwidth Improvements.} 
To further examine the performance impact of design levels and NAND devices, we measure the achievable read bandwidth because the bandwidth determines the inference performance. \reffigure\ref{fig:perf_tlc_slc_comp} shows the aggregate bandwidth comparison between high-density and low-latency SSDs for MoE and \llama models for channel-level and die-level designs. Dense inference can easily saturate channel-level or die-level peak internal bandwidth since the dense inference is characterized by highly sequential data access, which is suitable for NAND devices. However, the aggregate bandwidth decreases as the sparsity ratios increase. The bandwidth degradation is due to the fact that part of readout data are discarded when skipping non-activated neurons while the read latency for single page is fixed for both dense and sparse inferences. The other observation is that the sparse inference is better suited for large embedding sizes instead of small ones, \ie the bandwidth degradation of LLMs with larger embedding sizes is less. For example, DeepSeek-16B with smaller embedding size $dim_e=2048$ shows more significant performance degradation compared to other LLMs with embedding size $dim_e=4096$ or $5120$. 
As compared to external PCIe bandwidth, most channel-level designs outperform the 8GB/s of PCIe $4.0 \times 4$ while die-level designs with SLC exceed PCIe $4.0\times 16$ capacity (32 GB/s). The bandwidth benefits stem from: \Design exploits NSP to process data much faster internally than PCIe can transfer it, making local computation essential for avoiding PCIe bottlenecks.

\begin{figure}[t]
    \centering
    \includegraphics[width=0.75\linewidth]{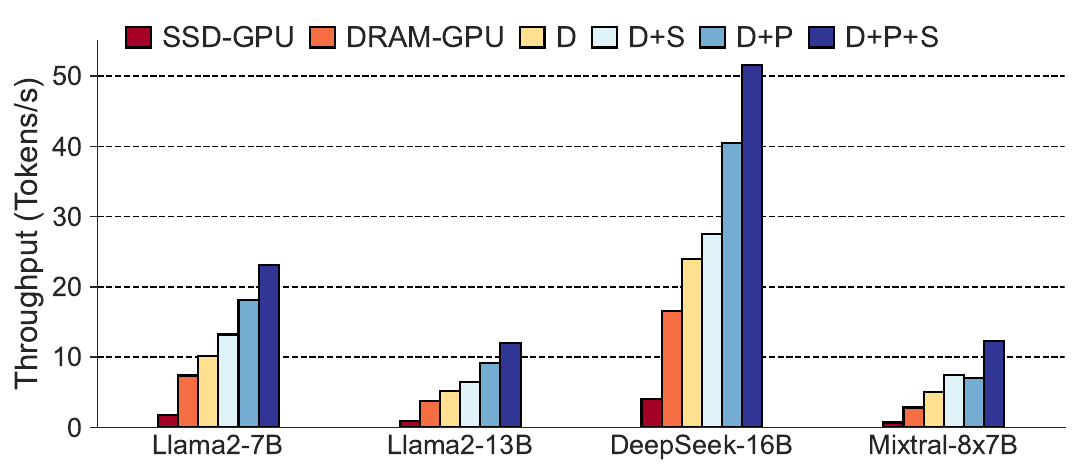}
    \caption{Impact of different optimization strategies (D=Die-level, S=Sparse with $50\%$ ratio, P=Pipeline) on the generation throughput (token/s). \Design uses the die-level design with SLC device.}
    \label{fig:perf_impact}
\end{figure}

\noindent\textbf{Impact of Optimization Schemes.} 
\reffigure~\ref{fig:perf_impact} summarizes the incremental impact of two proposed optimization schemes (sparse inference and pipelined execution scheme) on generation throughput (tokens/s). Here, \Design die-level design with SLC device is considered and compared to the GPU-centric solutions (SSD-GPU and DRAM-GPU designs). The GPU-centric accelerators fail to deliver satisfactory serving quality since their generation throughputs are too low to satisfy real-time interaction. In comparison, \Design die-level design with SLC without any optimizations consistently outperforms the two GPU-centric accelerators due to the higher internal bandwidth. 
After applying the sparse inference and pipelined execution, these two schemes jointly provide 1.8 to $2.4\times$ throughput improvements. The LLM sparsity skips non-activated neurons, thereby reducing the data amount of weight fetching. Then the pipelined execution scheme improves the hardware utilization of DRAM-PIM and SSD-NSP modules by overlapping consecutive layers. As a result, the generation throughput is effectively increased. \Design's die-level design with sparse inference and pipelined execution achieves up to $17.6\times$ performance improvement over SSD-GPU baseline.

\begin{figure}[t]
    \centering
    \begin{subfigure}{\linewidth}
        \centering
        \includegraphics[width=0.75\linewidth]{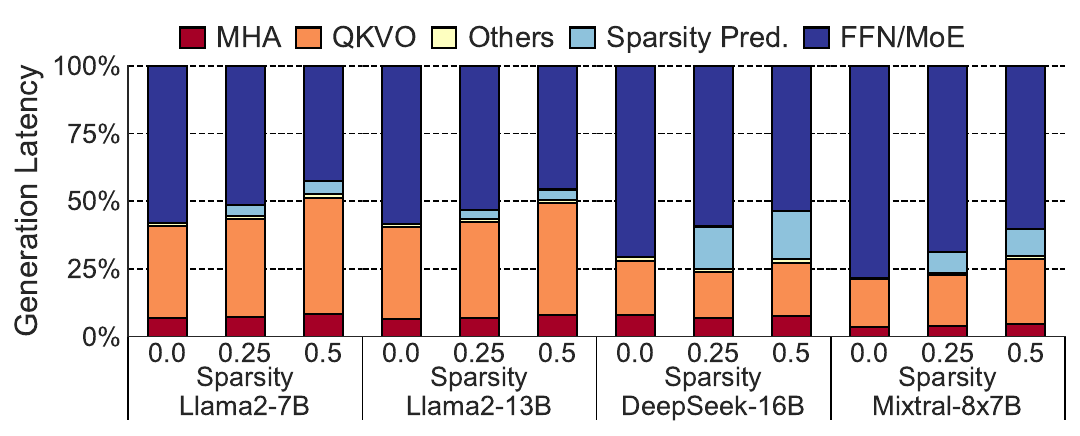}
        \caption{Latency breakdown}
    \end{subfigure}
    \begin{subfigure}{\linewidth}
        \centering
        \includegraphics[width=0.75\linewidth]{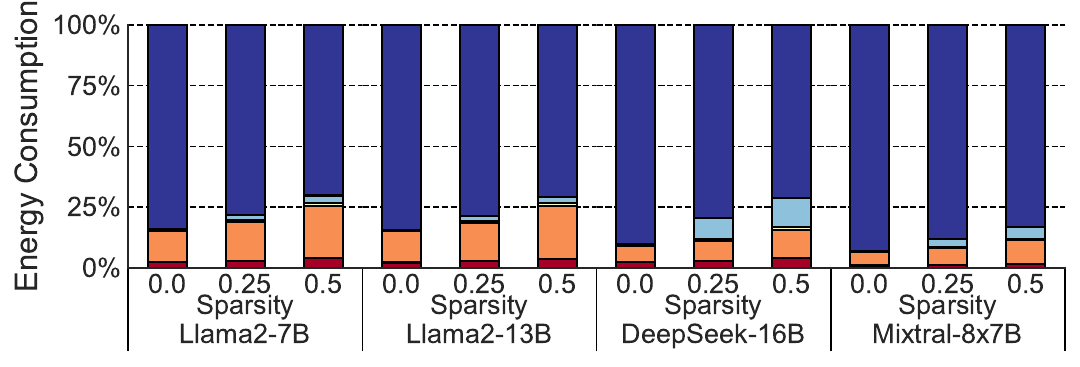}
        \caption{Energy breakdown}
    \end{subfigure}
    \caption{Performance breakdown results ((a) latency breakdown and (b) energy breakdown) for \Design die-level design with SLC NAND chips. The sparsity ratios are 0.0, 0.25, 0.5.}
    \label{fig:perf_breakdown}
\end{figure}

\subsection{Overhead Analysis}\label{subsec:analysis_error}
In this section, we perform the breakdown analysis to study the overhead of different components from both algorithm and hardware perspectives. The pipelined execution is disabled to provide complete runtime statistics.

\noindent\textbf{Runtime Breakdown Analysis.}
\reffigure\ref{fig:perf_breakdown} shows the generation latency breakdown and energy consumption breakdown of each module in \Design die-level design with sparsity prediction. The sparsity ratios vary from 0.0 to 0.5. 
QKVO and FFN/MoE layers contribute to over $80\%$ of total token generation latency and energy consumption during inference. 
The MoE-based LLMs, such as DeepSeek-16B and Mixtral-$8\times7$B, incur a higher overhead on processing FFN/MoE layers than non-MoE \llama. This is because the number of MoE weights in MoE-based models is much more than non-MoE models. The MHA is computed using DRAM's in-situ computing, which efficiently reduces data movement. Thus, the percentages of MHA runtime and energy consumption are relatively small ($<10\%$).

\noindent \textbf{Sparsity Prediction Overhead.}
The other type of overhead in \reffigure\ref{fig:perf_breakdown} is the sparsity prediction, which trades off additional low-rank sparsity prediction against the reduced FFN/MoE weight data fetching. As the sparsity ratios increase, the largest overhead due to FFN/MoE decreases. Note that the overhead of sparsity prediction in MoE-based LLMs is higher than non-MoE models since multiple sparsity predictions are required to run for all activated FFNs in the MoE module. DeepSeek-16B has the highest sparsity prediction overhead since its small embedding size $dim_e=2048$ leads to a more efficient inference that incurs less weight fetching. Other models except for DeepSeek-16B have $<10\%$ sparsity prediction runtime overhead.

\noindent \textbf{Area and Power Overhead.} 
Table~\ref{tab:hardware_specs} also summarizes the area and power of \Design near-storage accelerator. As mentioned, the ASIC systolic arrays are synthesized in TSMC 40nm technology and scaled to 22nm. The synthesized ASIC designs have 0.15mm$^2$ area overhead for channel-level PEs and 0.07mm$^2$ for die-level PEs. Compared to the SSD chip area~\cite{wang2024beacongnn}, the ASIC design increases less than $1\%$ of area overhead to the 3D NAND die for both low-latency and high-density devices. The power analysis also demonstrates a reasonable performance in tens of mW, which is comparable to the 3D NAND chip-level average power consumption of around 100mW~\cite{SSDPowerRef}. Combining with the DRAM active power of 3W per 8GB~\cite{JEDEC}, the total power consumption of \Design is estimated to be around 15W, which can be supported by PCIe 4.0 power delivery of 75W.

\subsection{Discussion}


\noindent\textbf{Implement Multi-plane Read.}
We anticipate that additional performance gain can be achieved through the use of multi-plane read operations~\cite{ONFI3.1}. Specifically, the fused FFN layer can be enhanced by strategically placing two consecutive physical pages on the same wordline or enabling multi-plane independent read operations~\cite{PIR}. These strategies are expected to improve the die-level NSP architecture.

\noindent\textbf{Cost Comparison.} 
Another advantage of \Design is its economic efficiency, characterized by high storage density and a lower cost-per-gigabyte of SSD compared to DRAM-GPU or SSD-GPU systems. Despite the high cost associated with GPU-based configurations, our design demonstrates competitive performance and up to 5 times higher power efficiency compared to the baselines. Furthermore, when considering only memory costs, the price of server DRAM is substantially higher—three orders of magnitude greater than that of TLC NAND~\cite{cost_comparison}.

\noindent\textbf{Advanced Interfaces.}
In comparison to our current experimental setup, the landscape of advanced technologies is expected to evolve significantly. For instance, PCIe 6.0~\cite{PCIe} and ONFI 6.0~\cite{ONFI6.0} are projected to support increased data transfer rates of up to 8GB/s per lane and 4.8GB/s per channel, respectively. Since the SSD interface speed growing faster than DRAM after advent of NVMe protocol, the gap between DRAM and storage is anticipated to narrow progressively. If such high throughput can support larger models within acceptable latency thresholds with the help of NSP, NSP may become increasingly viable for consideration.

\section{Conclusion}\label{sec:conclusion}
This work presents \Design, a novel heterogeneous DRAM-SSD accelerator that enables low-cost LLM inference serving for hardware-constraint local environments. We leverage LLM sparsity to address memory footprint and energy consumption issues. By combining NSP and PIM, \Design achieves substantial performance improvements: it outperforms existing SSD-GPU and DRAM-GPU systems, offering up to 13-18$\times$ throughput increase over SSD-GPU and up to 9-10$\times$ energy reduction over DRAM-GPU system. \Design design also incorporates that help to integrate existing SSD framework. This work paves the way for cost-effective LLM deployment in edge and local computing devices.

\begin{acks}
This work was supported in part by PRISM and CoCoSys, centers in JUMP 2.0, an SRC program sponsored by DARPA; SRC Global Research Collaboration (GRC) grant; and National Science Foundation (NSF) grants \#1826967, \#1911095, \#2052809, \#2112665, \#2112167, and \#2100237.
\end{acks}

\bibliographystyle{ACM-Reference-Format}
\bibliography{refs}

\end{document}